\documentclass[11pt,a4paper]{article}
\usepackage{amsmath,amsfonts,amssymb,amsthm}
\usepackage{fullpage,feynmp}
\newtheorem*{thm-ward}{Theorem \ref{thm:ward}}%[section]
\newtheorem{thm}{Theorem}%[section]

\newtheorem{defn}[thm]{Definition}
\newtheorem{ex}[thm]{Example}
\newtheorem{rem}[thm]{Remark}

\def\bar{\overline}
\def\C{\mathbb{C}}

\newcommand{\ins}[2]{#1~|~#2}

\def\F{\mathcal{F}}

\def\half{\tfrac{1}{2}}

\def\id{\mathrm{id}}

\def\L{\mathcal{L}}

\def\ren{\mathrm{ren}}

\def\res{\mathrm{res}}

\def\Sym{\mathrm{Sym}}

\def\tilde{\widetilde}

\def\vertphi{
{~
      \begin{fmfchar}(8,6)
	\fmfpen{.1mm}
	\fmfleft{l1,l2}
	\fmfright{r1,r2}
	\fmf{plain}{l1,v}
	\fmf{plain}{l2,v}
	\fmf{plain}{r1,v}
	\fmf{plain}{v,r2}
      \end{fmfchar}
}}

\def\el{
{~
    \begin{fmfgraph}(8,5)
      \fmfleft{l}
      \fmfright{r}
      \fmf{plain}{l,r}
    \end{fmfgraph}
  }
}

\def\ph{
{~
    \begin{fmfgraph}(8,5)
      \fmfleft{l}
      \fmfright{r}
      \fmf{photon}{l,r}
    \end{fmfgraph}
  }
}

\def\vertex{
{~
	\begin{fmfgraph}(8,5)
	  \fmfset{wiggly_len}{4pt} % reducing length of wiggles in boson lines
	  \fmfset{wiggly_slope}{70}
	  \fmfleft{l}
	  \fmfright{r1,r2}
	  \fmf{photon}{l,v}
	  \fmf{plain}{r1,v}
	  \fmf{plain}{v,r2}
	\end{fmfgraph}
      }
}

\def\qua{
{~
      \begin{fmfgraph}(8,6)
	\fmfpen{.1mm}
        \fmfforce{(0w,.15h)}{l}
        \fmfforce{(1w,.15h)}{r}
	\fmf{plain}{l,r}
      \end{fmfgraph}
}}

\def\gho{
{~
      \begin{fmfgraph}(8,6)
	\fmfpen{.1mm}
	\fmfset{dot_len}{.3mm}
        \fmfforce{(0w,.15h)}{l}
        \fmfforce{(1w,.15h)}{r}
        	\fmf{dots}{l,r}
    \end{fmfgraph}
}}

\def\glu{
{~
      \begin{fmfchar}(8,6)
	\fmfpen{.1mm}
	\fmfset{curly_len}{.5mm}
	\fmfleft{l}
	\fmfright{r}
	\fmf{gluon}{l,r}
      \end{fmfchar}
}}

\def\quaglu{
{~
      \begin{fmfchar}(8,6)
	\fmfpen{.1mm}
	\fmfset{curly_len}{.5mm}
	\fmfleft{l}
	\fmfright{r1,r2}
	\fmf{gluon}{l,v}
	\fmf{plain}{r1,v}
	\fmf{plain}{v,r2}
      \end{fmfchar}
}}

\def\ghoglu{
{~
      \begin{fmfchar}(8,6)
	\fmfpen{.1mm}
	\fmfset{curly_len}{.5mm}
	\fmfset{dot_len}{.3mm}
	\fmfleft{l}
	\fmfright{r1,r2}
	\fmf{gluon}{l,v}
	\fmf{dots}{r1,v}
	\fmf{dots}{v,r2}
      \end{fmfchar}
}}

\def\gluc{
{~
      \begin{fmfchar}(8,6)
	\fmfpen{.1mm}
	\fmfset{curly_len}{.5mm}
	\fmfleft{l}
	\fmfright{r1,r2}
	\fmf{gluon}{l,v}
	\fmf{gluon}{v,r1}
	\fmf{gluon}{v,r2}
      \end{fmfchar}
}}

\def\gluq{
{~
      \begin{fmfchar}(8,6)
	\fmfpen{.1mm}
	\fmfset{curly_len}{.5mm}
	\fmfleft{l1,l2}
	\fmfright{r1,r2}
	\fmf{gluon}{l1,v}
	\fmf{gluon}{l2,v}
	\fmf{gluon}{r1,v}
	\fmf{gluon}{v,r2}
      \end{fmfchar}
}}

\title{Multiplicative renormalization and Hopf algebras}
\author{
Walter D. van Suijlekom\\[5mm]
Institute for Mathematics, Astrophysics and Particle Physics\\
Faculty of Science, Radboud Universiteit Nijmegen\\
Toernooiveld 1, 6525 ED Nijmegen, The Netherlands\\
e-mail: \texttt{waltervs@math.ru.nl}}
\date{4 July 2007}

\begin{document}
\begin{fmffile}{graphs-green}

\fmfset{wiggly_len}{5pt} % reducing length of wiggles in boson lines
\fmfset{wiggly_slope}{70} % increasing slope of wiggles in boson lines
\fmfset{curly_len}{4pt} % reducing length of wiggles in boson lines
\fmfset{curly_len}{1.4mm}

\fmfset{dot_len}{1mm}

\maketitle

\begin{abstract}
We derive the existence of Hopf subalgebras generated by Green's functions in the Hopf algebra of Feynman graphs of a quantum field theory. 
This means that the coproduct closes on these Green's functions. It allows us for example to derive Dyson's formulas in quantum electrodynamics relating the renormalized and bare proper functions via the renormalization constants and the analogous formulas for non-abelian gauge theories. In the latter case, we observe the crucial role played by Slavnov--Taylor identities. 
\end{abstract}

\section{Introduction}
During the last decade much of the combinatorial structure of renormalization of perturbative quantum field theories has been understood in terms of Hopf algebras (starting with \cite{Kre98,CK99}). Although this led to many insights in the process of renormalization one could argue that since the elements in the Hopf algebra are individual Feynman graphs, it is rather unphysical. Rather, one would like to describe the renormalization process on the level of the 1PI Green's functions. Especially for (non-abelian) gauge theories, the graph-by-graph approach of for instance the BPHZ-procedure is usually replaced by more powerful methods based on BRST-symmetry and the Zinn-Justin equation (and its far reaching generalization: the Batalin-Vilkovisky formalism). They all involve the 1PI Green's functions or even the full effective action that is generated by them. 

The drawback of these latter methods, is that they rely heavily on functional integrals and are therefore completely formal. The good thing about BPHZ-renormalization was that if one accepts the perturbative series of Green's function in terms of Feynman graphs as a starting point, the procedure is completely rigorous. Of course, this allowed the procedure to be described by a mathematical structure such as a Hopf algebra. 

In this article, we address the question whether we can prove some of the results on Green's functions starting with the Hopf algebra of Feynman graphs. We derive the existence of Hopf subalgebras generated by the 1PI Green's functions. We do this by showing that the coproduct takes a closed form on these Green's functions, thereby relying heavily on a formula that we have previously derived.

In \cite{BK05} Hopf subalgebras were given for any connected graded Hopf algebra as solutions to Dyson-Schwinger equations. It turned out that there was a close relation with Hochschild cohomology. 
For quantum electrodynamics, certain Hopf subalgebras of planar binary tree expansions  were considered in \cite{BF00} (cf. also \cite{BF03}). Via a noncommutative Hopf algebra of formal diffeomorphisms (see also \cite{BFK06}), the authors derived Dyson's formulas relating the renormalized and unrenormalized proper functions. 
The case of non-abelian gauge theories was discussed by Kreimer in \cite{Kre05,Kre06} where it was claimed that the existence of Hopf subalgebras follows from the validity of the Slavnov--Taylor identities {\it inside} the Hopf algebra of (QCD) Feynman graphs. We now fully prove this claim by applying a formula for the coproduct on Green's functions that we have derived before in \cite{Sui07}. In fact, that formula allowed us to prove compatibility of the Slavnov--Taylor identities with the Hopf algebra structure.

After recalling the preliminaries on Feynman graphs, Green's functions and some combinatorial factors, we state the key formula for the coproduct on Green's functions. We first consider a scalar field theory ($\phi^4$) and derive the well-known relations between the renormalized and bare proper functions and the renormalization constants (Eq. \eqref{green-scalar} below). 

Then, we consider quantum electrodynamics, for which we derive Dyson's formulas \cite{Dys49}:
$$
\Gamma^\mu_\ren (e) = Z_2 Z_3^{1/2} \Gamma^\mu (e_0),\qquad
\Sigma_\ren(e) = Z_2 \Sigma(e_0), \qquad
\Pi_\ren(e) = Z_3 \Pi(e_0),
$$
with $e_0$ the bare electric charge and $e$ the renormalized charge (this is Eq. \eqref{dyson} below).

Finally, we establish a Hopf subalgebra consisting of the Green's functions in a non-abelian gauge theory. Here the Slavnov--Taylor identities turn out to play a crucial role. Again, we have the well-known formulas for the renormalized and bare functions (Eq. \eqref{green-qcd} below). 

Note that these formulas for the proper functions are not derived via the usual procedure of adding counterterms to the Lagrangian but follow from the Hopf algebraic structure in combination with the Fyenman rules.

\section{Hopf algebra of Green's functions}
We will prove the existence of a Hopf subalgebra in $H$ generated by the three 1PI Green's functions relevant for renormalization of quantum field theories. In particular, we will consider Hopf subalgebras in the case of $\phi^4$-theory, quantum electrodynamics (QED) and quantum chromodynamics (QCD).
We start by briefly recalling the relevant definitions and results from \cite{Sui07} while referring the reader to that paper for more details.
\subsection{Preliminaries}
Our starting point is a renormalizable quantum field theory, given for instance by a Lagrangian $\L$. In perturbation theory, one usually associates to each term in the Lagrangian an edge or a vertex and starts to built Feynman diagrams from them. It is well-known that for the purpose of renormalization theory, it is enough to consider only one-particle irreducible (1PI) diagrams with external structure corresponding to each term in the Lagrangian. For example, in $\phi^4$-theory, there is one vertex of valence 4 and one edge, and we consider only diagrams with 2 and 4 external edges. In general, we will consider sums over all 1PI diagrams with the same external structure and this defines the 1PI Green's functions
\begin{gather*}
G^v = 1 + \sum_{\res(\Gamma)=v} \frac{\Gamma}{\Sym(\Gamma)}, \qquad
G^e = 1 - \sum_{\res(\Gamma)=e} \frac{\Gamma}{\Sym(\Gamma)}.
\end{gather*}
with $v$ a vertex and $e$ and edge. 
Here $\res(\Gamma)$ is the {\it residue} of $\Gamma$ ({i.e.} the vertex/edge the graph $\Gamma$ corresponds to after collapsing all internal points), and the symmetry factor $\Sym(\Gamma)$ is the order of the automorphism group of the graph. It is extended to disjoint unions of graphs by setting 
\begin{equation*}
\Sym(\Gamma \cup \Gamma') =  \left(n(\Gamma, \Gamma')+1 \right) \Sym(\Gamma) \Sym(\Gamma'),
\end{equation*}
with $n(\Gamma, \Gamma')$ the number of connected components of $\Gamma$ that are isomorphic to $\Gamma'$.

\bigskip

In \cite{CK99}, Connes and Kreimer defined a coproduct on Feynman diagrams. This encodes the procedure of renormalization in terms of a Hopf algebra (see the appendix for a quick overview of Hopf algebras). Let us briefly recall how the coproduct was defined. One considers the algebra $H$ generated by Feynman diagrams (for some quantum field theory), on which a coproduct $\Delta: H \to H \otimes H$ is defined by
\begin{equation}
\label{cop}
\Delta(\Gamma) = \Gamma \otimes 1 + 1 \otimes \Gamma + \sum_{\gamma \subsetneq \Gamma} \gamma \otimes \Gamma
\end{equation}
where the sum is over all subdiagrams $\gamma$ that are disjoint unions of 1PI diagrams. Moreover, a counit $\epsilon : H \to \C$ is defined as the algebra map that takes the value 1 on the identity and zero on any 1PI graph. The antipode $S: H \to H$ can be defined recursively by 
\begin{equation*}
S(\Gamma) = - \Gamma - \sum_{\gamma \subsetneq \Gamma} S(\gamma) \Gamma/\gamma.
\end{equation*}
It turned out that the BPHZ-procedure of recursively subtracting the divergent part of a Feynman amplitude $U(\Gamma)$ (for a given graph $\Gamma$) in order to give the renormalized amplitude $R(\Gamma)$, is given by a convolution product in the space of maps from $H$ to some space of functions depending on the regularization parameter. Indeed, the Feynman amplitude $U$ can be understood as such a map: $\Gamma \to U(\Gamma)$. The counterterms are given by the map $C$,
\begin{align}
\label{counterterm}
C(X) = \epsilon(X) - T \left[ (C \otimes U ) \big( (\id \otimes (1-\epsilon) )\Delta (X) \big) \right]
\end{align}
with $T$ a map that projects onto the part of the amplitude that diverges when the regularization parameter goes to 0 (or infinity in the case of a cutoff). Crucial in proving that $C$ is an algebra map is the following multiplicative property $T(XY) = T(T(X)Y) + T(X T(Y)) - T(X)T(Y)$, which motivated the study of so-called Rota-Baxter algebras within the context of renormalization (see \cite{EG07} and references therein). 
In the case of dimensional regularization, the regularizing parameter is the complex number $z$ (working in $d-z$ dimensions) and $T$ is the projection onto the pole part of the Laurent series in $z$ which indeed satisfies the multiplicative property. 
The renormalized Feynman amplitude $R$ is given as the convolution product:
\begin{equation*}
R(X)=(C \ast U) (X) := (C \otimes U)\left(\Delta(X)\right).
\end{equation*}
\begin{rem}
That this indeed encodes the BPHZ-procedure can be seen as follows. Let $\Gamma$ be a 1PI graph. Then, with the coproduct given by \eqref{cop} we obtain
\begin{align*}
C(\Gamma) &= -T \left[ U(\Gamma) + \sum_{\gamma \subsetneq \Gamma} C(\gamma) U(\Gamma/\gamma) \right ] = -T \left[ \bar R ( \Gamma ) \right] \\
R(\Gamma) &= \bar R(\Gamma) + C(\Gamma)
\end{align*}
where $\bar R$ is the so-called {\it prepared amplitude}. See \cite[Sect. 5.3.2]{Col84} for more details on the BPHZ-procedure.
\end{rem}

In the next sections, we would like to derive a closed form of the coproduct on Green's functions. We do this using a formula that we have derived in \cite{Sui07} and have shown to imply compatibility of the above coproduct with Ward identities in quantum electrodynamics and Slavnov--Taylor identities in non-abelian gauge theories. It turned out that the corresponding Hopf algebras can be consistently quotiented by these identities, giving Hopf algebras that have them `built in'. From this, one can deduce that if the unrenormalized Feynman amplitudes satisfy the Ward or Slavnov-Taylor identities, then so do the renormalized ones as well as the counterterms. 

Before stating the aforementioned formula, we introduce some notation. Let $L(\Gamma)$ denote the number of loops of $\Gamma$ and $\ins{\Gamma}{\gamma}$ the number of ways to insert $\gamma$ inside $\Gamma$. Explicitly, the latter is given by 
\begin{align}
\label{ins-places}
\ins{\Gamma}{\gamma} 
&=\prod_i n_{v_i}(\gamma)! {V_i(\Gamma)  \choose n_{v_i}(\gamma)} \prod_{j} n_{e_j}(\gamma)! {I_j(\Gamma) + n_{e_j}(\gamma)-1  \choose n_{e_j}(\gamma)}.
\end{align}
Here $V_i(\Gamma)$ is the number of vertices in $\Gamma$ of type $i$ and $I_j(\Gamma)$ the number of internal edges in $\Gamma$ of type $j$. Moreover, 
$n_r(\gamma)$ is the number of connected components of $\gamma$ with residue $r$ ($r$ being a vertex or an edge). Indeed, then the binomial coefficients arises for each vertex $v_i$ since we are choosing $n_{v_i}$ out of $V_i$ whereas for an edge $e_j$ we choose $n_{e_j}$ out of $I_j$ {\it with repetition} because of multiple insertions of self-energy graphs on the same edge of $\Gamma$. See for more details \cite{Sui07}, where we have also derived the key formula for the coproduct on the 1PI Green's functions:
\begin{equation}
\label{cop-green}
\Delta \left(G^r \right)=  \sum_{\gamma} \sum_{\res(\Gamma)=r} 
\frac{\ins{\Gamma}{\gamma}}{\Sym(\gamma)\Sym(\Gamma)}  ~ \gamma \otimes \Gamma.
\end{equation}
The sum is over all $\gamma$ which are disjoint unions of 1PI graphs, whereas $\Gamma$ is 1PI with the indicated residue $r$. Note that this formula holds for the 1PI Green's functions for any quantum field theory, by simply allowing $r$ to be vertices and edges of different types (photon, electron, gluon, etc...). 

\begin{ex}
We illustrate the combinatorial factors introduced above with the following example. Consider the graph 
$$
\Gamma :=\parbox{30pt}{\begin{fmfgraph*}(30,40)
  \fmfleft{l}
  \fmfright{r}
  \fmf{photon}{l,v,r}
  \fmfv{decor.shape=circle, decor.filled=0, decor.size=5thick}{v}
  \end{fmfgraph*}}.
$$
Then $\res(\Gamma) = \ph$ and $Sym(\Gamma)=2$. Moreover, for the number of insertion places, we have for instance:
\begin{align*}
\parbox{30pt}{
  \begin{fmfgraph*}(30,40)
  \fmfleft{l}
  \fmfright{r}
  \fmf{photon}{l,v,r}
  \fmfv{decor.shape=circle, decor.filled=0, decor.size=5thick}{v}
  \end{fmfgraph*}
} 
~\Big|~
\parbox{40pt}{
\begin{fmfgraph*}(40,40)
      \fmfleft{l}
      \fmfright{r1,r2}
      \fmf{photon}{l,v}
      \fmf{plain}{v,v1,r1}
      \fmf{plain}{v,v2,r2}
      \fmffreeze
      \fmf{photon}{v1,v2}
\end{fmfgraph*}
} ~ = {2 \choose 1} = 2
\quad \text{ whereas }\quad 
\parbox{30pt}{
  \begin{fmfgraph*}(30,40)
  \fmfleft{l}
  \fmfright{r}
  \fmf{photon}{l,v,r}
  \fmfv{decor.shape=circle, decor.filled=0, decor.size=5thick}{v}
  \end{fmfgraph*}
} 
~\Big|~
\parbox{30pt}{
\begin{fmfgraph*}(30,30)
      \fmfleft{l}
      \fmfright{r}
      \fmf{plain}{l,v1,v2,r}
      \fmf{photon,left,tension=0}{v1,v2}
\end{fmfgraph*}
} ~ 
\parbox{30pt}{
\begin{fmfgraph*}(30,30)
      \fmfleft{l}
      \fmfright{r}
      \fmf{plain}{l,v1,v2,v5,v6,r}
      \fmf{photon,left,tension=0}{v1,v5}
      \fmf{photon,right,tension=0}{v2,v6}
\end{fmfgraph*}
} ~ = 2! {3 \choose 2}  = 6.
\end{align*}
\end{ex}

\subsection{Scalar field theory}
As a warming-up for the next sections where we consider QED and QCD, we consider $\phi^4$-theory. The Feynman diagrams are constructed from one type of vertex (of valence 4) and one type of edge. As mentioned above, we will consider only the 2 and 4 points Green's functions $G^{(2)} := G^\el$ and $G^{(4)} := G^\vertphi$. 

We start by simplifying the expression $\ins{\Gamma}{\gamma}$ a bit. Recall that $n_v(\gamma)$ and $n_e(\gamma)$ denote the numbers that count the vertex and self-energy graphs in $\gamma$, respectively. From the presence of only one vertex in the theory, one easily obtains the two formulas $L=I-(V-1)$ and $4V=E+2I$, relating the number of internal lines $I$, external lines $E$ and the number of vertices $V$ of $\Gamma$ to the loop number $L=L(\Gamma)$. Inserting the resulting expressions for $I$ and $V$ (in terms of $L$ and $E$) into Eq. \eqref{ins-places} yields
\begin{align*}
\ins{\Gamma}{\gamma} &= 
\left\{ \begin{array}{ll}
n_v(\gamma) ! { L \choose n_v(\gamma)} n_e(\gamma)! { 2L + n_e(\gamma) -2 \choose n_e(\gamma) } &\text{ if } \Gamma \text{ is a vertex graph,}
\\[4mm]
n_v(\gamma) ! { L+1 \choose n_v(\gamma)} n_e(\gamma)! { 2L + n_e(\gamma) -1 \choose n_e(\gamma) } &\text{ if } \Gamma \text{ is a self-energy graph.}
\end{array} \right.
\end{align*}
Let us now consider the coproduct on the two Green's functions $G^{(2)}$ and $G^{(4)}$ by inserting these expressions in Eq. \eqref{cop-green}. Clearly, we can split the sum over $\gamma$ in two parts: $\gamma_V$ and $\gamma_E$ containing only vertex and self-energy graphs respectively. This gives
\begin{align*}
	\Delta(G^{(2)})&= \sum_{L=0}^\infty \sum_{\begin{smallmatrix} E(\Gamma)=2 \\ L(\Gamma)=L \end{smallmatrix}} 
\left[ \sum_{\gamma_V} n_V ! { L \choose n_V } \frac{\gamma_V}{\Sym(\gamma_V)} \right]
\left[\sum_{\gamma_E} n_E ! { 2 L  + n_E -2 \choose n_E } \frac{\gamma_E}{\Sym(\gamma_E)} \right]
\otimes \frac{\Gamma}{\Sym(\Gamma)},\\
	\Delta (G^{(4)})&= \sum_{L=0}^\infty \sum_{\begin{smallmatrix} E(\Gamma)=4 \\ L(\Gamma)=L \end{smallmatrix}} 
\hspace{-1mm}
\left[ \sum_{\gamma_V} n_V ! { L+1 \choose n_V } \frac{\gamma_V}{\Sym(\gamma_V)} \right]
\hspace{-1mm}
\left[\sum_{\gamma_E} n_E ! { 2 L  + n_E -1 \choose n_E } \frac{\gamma_E}{\Sym(\gamma_E)} \right]
\hspace{-1mm}
\otimes 
\frac{\Gamma}{\Sym(\Gamma)},
\end{align*} 
where we have used the shorthand notation $n_V := n_v(\gamma_V)$ and $n_E := n_e(\gamma_E)$.
We will now evaluate each of the sums between square brackets. First, let us
fix $n_V$ and restrict the sum over $\gamma_V$ to graphs consisting of $n_V$ 1PI graphs. Then,\footnote{We use the notation $h^0(\gamma)$ for the number of connected components of a graph $\gamma$, in accordance with the usual notation for the Betti numbers.}
\begin{align*}
 \sum_{h^0(\gamma_V)=n_V} n_V ! { L \choose n_V } \frac{\gamma_V}{\Sym(\gamma_V)} 
&= \sum_{h^0(\gamma_V)=n_V} 
\left( 
\sum_{\begin{smallmatrix} \gamma_v,\tilde\gamma_V \\ \gamma_v \tilde \gamma_V \simeq \gamma_V \end{smallmatrix}}  
\frac{n(\tilde \gamma_V, \gamma_v) + 1}{n_V} \right) 
n_V! { L \choose n_V } \frac{\gamma_V}{\Sym(\gamma_v)},
\end{align*}
where we have simply inserted 1. Indeed, for fixed $\gamma_V$ we have
\begin{align*}
\sum_{\begin{smallmatrix} \gamma_v,\tilde\gamma_V \\ \gamma_v \tilde \gamma_V \simeq \gamma_V \end{smallmatrix}}  
\frac{n(\tilde \gamma_V, \gamma_v) + 1}{n_V} = \sum_{\gamma_v} \frac{n(\gamma_V,\gamma_v) }{n_V} = 1.
\end{align*}
A glance back at the definition of $\Sym(\gamma_v \tilde\gamma_V)$ yields for the above sum
\begin{align*}
\sum_{\gamma_v}  \frac{\gamma_v}{\Sym(\gamma_v)} 
\sum_{h^0(\tilde\gamma_V)=n_V-1} 
\hspace{-3mm}
(n_V-1)! { L \choose n_V } \frac{\tilde\gamma_V}{\Sym(\tilde\gamma_V)}
&= (G^{(4)} -1) 
\hspace{-3mm}
\sum_{h^0(\tilde\gamma_V)=n_V-1} 
\hspace{-3mm}
(n_V-1)! { L \choose n_V } \frac{\tilde\gamma_V}{\Sym(\tilde\gamma_V)}.
\end{align*}
Iterating this argument $n_V$ times and summing over $n_V$ gives
\begin{align*}
\sum_{\gamma_V} n_V ! { L \choose n_V } \frac{\gamma_V}{\Sym(\gamma_V)}
= \sum_{n_V=0}^\infty { L \choose n_V } \left(G^{(4)} -1 \right)^{n_V} 
= \left( G^{(4)} \right)^L
\end{align*}
Similarly, we derive
\begin{align*}
\sum_{\gamma_E} n_E ! { 2 L  + n_E -2 \choose n_E } \frac{\gamma_E}{\Sym(\gamma_E)} = \sum_{n_E=0}^\infty { 2L+ n_E -2 \choose n_E } \left(1- G^{(2)} \right)^{n_E} = \frac{1}{\left( G^{(2)} \right)^{2L-1} },
\end{align*}
from which we obtain
\begin{align}
\label{cop-2point}
\Delta \left( G^{(2)} \right) = \sum_{L=0}^\infty \frac{\left( G^{(4)} \right)^L } {\left( G^{(2)} \right)^{2L-1} } \otimes G^{(2)}_L.  
\end{align}
In like manner, one can show
\begin{align}
\label{cop-4point}
\Delta \left( G^{(4)} \right) = \sum_{L=0}^\infty \frac{\left( G^{(4)} \right)^{L+1}} {\left( G^{(2)} \right)^{2L} } \otimes G^{(4)}_L.  
\end{align}
In other words, the coproduct closes on the proper 2 and 4-point functions in $\phi^4$ and hence they form a Hopf subalgebra.

\subsubsection{Renormalized amplitudes and counterterms}
Via the Feynman rules one obtains the (unrenormalized) amplitudes -- denoted $U(\Gamma)$ -- for a graph $\Gamma$. 
By summing over all vertex and self-energy graphs and extending $U$ by linearity, one defines the unrenormalized proper 2 and 4-point functions (in the presence of a regularization) by
\begin{align*}
\Gamma^{(n)}(\lambda) &= U \left(G^{(n)} \right), \qquad n=2,4.
\end{align*}
We have explicitly denoted the dependence on the coupling constant $\lambda$ (as present in the original unrenormalized Lagrangian), but ignored for simplicity the momenta that are put on the external legs. The renormalized proper 2 and 4-point functions are given by 
\begin{align*}
\Gamma_\ren^{(n)}(\lambda) &= R \left(G^{(n)} \right), \qquad n=2,4.
\end{align*}
Finally, the renormalization constants $Z_1$ and $Z_2$ are defined by
\begin{align*}
Z_1= C  \left(G^{(4)} \right),\qquad
Z_2= C  \left(G^{(2)} \right).
\end{align*}
Recall that the maps $R,C$ and $U$ are related by the convolution product: $R=C \ast U$. In combination with Equation \eqref{cop-2point} this implies the following relation between the unrenormalized and renormalized proper functions and the counterterms:
\begin{align*}
\Gamma_\ren^{(2)} (\lambda) = \left( C \ast U \right) \left(G^{(2)}\right) 
= \sum_{L=0}^\infty \frac{Z_1^L}{Z_2^{2L-1}} \Gamma_L^{(2)} (\lambda)
= Z_2 \Gamma^{(2)}(\lambda_0),
\end{align*}
where $\lambda_0 := \frac{Z_1 \lambda}{Z_2^2}$ is the bare coupling constant. Indeed, $\Gamma_L^{(2)}$ contains $L$ powers of $\lambda$, one for each vertex. A similar computation can be done for the proper 4-point function, leading to the well-known relations (cf. for instance Equation (8-100) in \cite{IZ1980})
\begin{align}
\label{green-scalar}
\Gamma_\ren^{(n)} (\lambda)= Z_2^{n/2} \Gamma^{(n)}(\lambda_0); \qquad (n=2,4).
\end{align}

\subsection{Quantum electrodynamics}
The three Green's functions that are of interest in renormalization of quantum electrodynamics are $G^\vertex, G^\el$ and $G^\ph$ and correspond to the vertex, electron edge and photon edge. As in the previous subsection, we would like to establish that the coproduct has a closed form on these Green's functions. Let us start again by simplifying the expressions $\ins{\Gamma}{\gamma}$ that appear in Eq. \eqref{cop-green}. Since also in QED there is only one vertex, one can derive the following equalities for the number of vertices and electron and proton edges in a graph $\Gamma$ at loop order $L$:
\begin{equation}
\label{relations-qed}
\begin{aligned}
V &= 2 L + E_e+E_p - 2;\\
I_e &= 2 L +  \half E_e+E_p - 2,\\
I_p &= L + \half E_e  - 1.
\end{aligned}
\end{equation}
If we insert these expressions in Eq. \eqref{ins-places}, then Eq. \eqref{cop-green} implies
\begin{align*}
\Delta \left( G^\vertex \right)&= \sum_{L=0}^\infty       
\left[ \sum_{\gamma_V} n_V ! { 2L+1 \choose n_V } \frac{\gamma_V}{\Sym(\gamma_V} \right]
\left[ \sum_{\gamma_E} n_E ! { 2L+n_E - 1 \choose n_E } \frac{\gamma_V}{\Sym(\gamma_E} \right] \\ 
& \hspace{7cm} \times \left[ \sum_{\gamma_P} n_P ! { L+n_P - 1 \choose n_P } \frac{\gamma_V}{\Sym(\gamma_P} \right]
\otimes G^\vertex_L \\ 
\end{align*}
\begin{align*}
\Delta \left( G^\el \right) &=  \sum_{L=0}^\infty       
\left[ \sum_{\gamma_V} n_V ! { 2L \choose n_V } \frac{\gamma_V}{\Sym(\gamma_V} \right]
\left[ \sum_{\gamma_E} n_E ! { 2L+n_E - 2 \choose n_E } \frac{\gamma_E}{\Sym(\gamma_E} \right] \\ 
& \hspace{7cm} \times
\left[ \sum_{\gamma_P} n_P ! { L+n_P - 1 \choose n_P } \frac{\gamma_P}{\Sym(\gamma_P} \right]
\otimes G^\el_L \\
\Delta \left( G^\ph \right) &= \sum_{L=0}^\infty       
\left[ \sum_{\gamma_V} n_V ! { 2L \choose n_V } \frac{\gamma_V}{\Sym(\gamma_V} \right]
\left[ \sum_{\gamma_E} n_E ! { 2L+n_E - 1 \choose n_E }  \frac{\gamma_E}{\Sym(\gamma_E}\right] \\ 
& \hspace{7cm} \times
\left[ \sum_{\gamma_P} n_P ! { L+n_P - 2 \choose n_P } \frac{\gamma_P}{\Sym(\gamma_P} \right]
\otimes G^\ph_L .
\end{align*}
where $n_V := n_v(\gamma_V), n_E ;= n_e(\gamma_E)$ and $n_P:=n_p(\gamma_P)$ for vertex, electron self-energy and vacuum polarization graphs.

A computation very similar to that of the previous section allows one to rewrite the terms in brackets as powers of $G^\vertex, G^\el$ and $G^\ph$. Explicitly, we obtain for the three 1PI Green's functions:
\begin{equation}
\label{cop-qed}
\begin{aligned}
\Delta\left( G^\vertex \right) &= \sum_{L=0}^\infty 
\frac{\left( G^\vertex \right)^{2L+1} }{ \left( G^\el \right)^{2L} \left( G^\ph \right)^{L} } \otimes G^\vertex
\\
\Delta\left( G^\el \right)&= \sum_{L=0}^\infty
\frac{\left( G^\vertex \right)^{2L} }{ \left( G^\el \right)^{2L-1} \left( G^\ph \right)^{L} } \otimes G^\el
\\
\Delta\left( G^\ph \right)&= \sum_{L=0}^\infty
\frac{\left( G^\vertex \right)^{2L} }{ \left( G^\el \right)^{2L} \left( G^\ph \right)^{L-1} } \otimes G^\ph
\end{aligned}
\end{equation}
Hence, also in the case of quantum electrodynamics the coproduct closes on the 1PI Green's functions thereby generating a Hopf subalgebra. 

\begin{rem}
\label{rem:hopf-ideal}
From these formulas, the mentioned compatibility of the coproduct with the Ward identities $G^\vertex = G^\el$ is now an easy consequence. Indeed, 
\begin{multline*}
\Delta \left(G^\vertex - G^\el \right)
= \sum_{L=0}^\infty \frac{\left( G^\vertex \right)^{2L} }{ \left( G^\el \right)^{2L-1} \left( G^\ph \right)^{L} } \otimes \left[G^\vertex_L - G^\ph_L \right] \\
+ \sum_{L=0}^\infty  \left[G^\vertex - G^\ph \right]  \frac{\left( G^\vertex \right)^{2L} }{ \left( G^\el \right)^{2L} \left( G^\ph \right)^{L} } \otimes G^\vertex_L,
\end{multline*}
from which it follows at once that the ideal $I$ generated by $G^\vertex_L - G^\ph_L$ ($L=1,2,\cdots$) is a {\it Hopf ideal} (see the appendix). Consequently, the Hopf algebra $H$ can be quotiented by $I$ to give again a Hopf algebra $\tilde H$ which has the Ward identities built in.
\end{rem}

\subsubsection{Dyson's formula}
In \cite{Dys49} Dyson derived formulas relating the unrenormalized and renormalized proper functions and counterterms for quantum electrodynamics; they are the analogue of Eq. \eqref{green-scalar} above. In this section, we will derive them using the above closed form of the coproduct on the 1PI Green's functions, while never referring to the Lagrangian.

As before, the Feynman rules give rise to amplitudes $U(\Gamma)$ for each QED Feynman diagram $\Gamma$. At the level of Green's functions, we define the unrenormalized proper vertex function, electron self-energy and vacuum polarization by the identities (adopting also the notation that is common in the physics literature):
\begin{align*}
\Gamma^\mu (e) = U \left( G^\vertex \right),\qquad
\Sigma (e) = U \left( G^\el \right),\qquad
\Pi^{\mu\nu} (e) = U \left( G^\ph \right).	
\end{align*}
We have explicitly indicated the dependence on the electric charge $e$, but ignored for simplicity the momenta that are put on the external legs.
The renormalized proper functions $\Gamma^\mu_\ren(e), \Sigma_\ren(e)$ and $\Pi^{\mu\nu}_\ren(e)$  are defined by replacing $U$ by $R$ in the above formulas. Finally, the three corresponding renormalization constants are defined by
\begin{gather*}
Z_1 = C\left(G^\vertex \right), \qquad
Z_2 = C\left(G^\el \right), \qquad
Z_3 = C\left(G^\ph \right).
\end{gather*}
Dyson's formulas can now easily be derived by applying $R= C \ast U$ to $G^\vertex,G^\el$ and $G^\ph$ thereby using Eq. \eqref{cop-qed}. Recall that the bare electric charge $e_0$ is related to $e$ via the usual formula: $e_0 = \frac{Z_1 e} {Z_2 Z_3^{1/2}}$. 
A simple counting of the powers of $e$ ({i.e.} the number of vertices) in the proper functions at loop order $L$ then gives
\begin{equation}
\begin{aligned}
\label{dyson}
\Gamma^\mu_\ren (e) &= \sum_{L=0}^\infty 
\frac{Z_1^{2L+1} }{ Z_2^{2L} Z_3^{L} }~  \Gamma^\mu_L(e) = 
Z_2 Z_3^{1/2} \Gamma^\mu (e_0),
\\
\Sigma_\ren(e) &= \sum_{L=0}^\infty
\frac{Z_1^{2L} }{ Z_2^{2L-1} Z_3^{L} } \Sigma_L(e) =
Z_2 \Sigma(e_0),
\\
\Pi^{\mu\nu}_\ren(e) &= \sum_{L=0}^\infty
\frac{Z_1^{2L} }{ Z_2^{2L} Z_3^{L-1} } \Pi^{\mu\nu}_{\mu\nu,L} (e) =
Z_3 \Pi^{\mu\nu}(e_0).
\end{aligned}
\end{equation}

\begin{rem}
Let us come back once more to the Ward identities. Suppose we have chosen a regularization which respects the Ward identities, so that $U$ satisfies them in the physical sense:\footnote{Again we ignore the external momenta for the sake of simplicity; that this can be done was in fact shown in \cite{Sui06}.}
$$
\Gamma^\mu - \Sigma = 0.
$$
As a consequence, $U$ vanishes on the ideal $I$ (since it is generated by $G^\vertex - G^\el$) and is thus defined on the quotient $\tilde H = H/I$. Now, since $I$ is a Hopf ideal (cf. Remark \ref{rem:hopf-ideal}) it is shown in \cite{CK99} that $C$ is again a map of Hopf algebras so that $C$ vanishes on $I$ as well. Since $R = C \ast U$, we also have that $R(I)=0$ so that both the renormalized proper functions as well as the counterterms satisfy the Ward identities, leading in particular to the well-known expression $Z_1 = Z_2$ \cite{War50}.
\end{rem}

\subsection{Quantum chromodynamics}
We next consider the case of a non-abelian gauge theory. In order to be as concrete as possible, we consider quantum chromodynamics. There are the quark, ghost and gluon propagators, denoted
\begin{align*}
\raisebox{-7.5pt}{
\parbox{30pt}{
  \begin{fmfgraph}(30,10)
      \fmfleft{l}
      \fmflabel{}{l}
      \fmfright{r}
      \fmf{plain}{l,r}
  \end{fmfgraph}
}}
~,\qquad
\raisebox{-7.5pt}{
\parbox{30pt}{
  \begin{fmfgraph}(30,10)
      \fmfleft{l}
      \fmflabel{}{l}
      \fmfright{r}
      \fmf{dots}{l,r}
  \end{fmfgraph}
}}
~,\qquad
\raisebox{-7.5pt}
{
\parbox{30pt}{
  \begin{fmfgraph}(30,10)
      \fmfleft{l}
      \fmflabel{}{l}
      \fmfright{r}
      \fmf{gluon}{l,r}
  \end{fmfgraph}
}}~,
\end{align*}
respectively, and four vertices: 
\begin{align*}
\raisebox{-7.5pt}{
\parbox{30pt}{
    \begin{fmfchar}(30,20)
      \fmfleft{l}
      \fmfright{r1,r2}
      \fmf{gluon}{l,v}
      \fmf{plain}{r1,v}
      \fmf{plain}{v,r2}
    \end{fmfchar}
  }}
~,\qquad
\raisebox{-7.5pt}{
\parbox{30pt}{
  \begin{fmfgraph}(30,20)
      \fmfleft{l}
      \fmfright{r1,r2}
      \fmf{gluon}{l,v}
      \fmf{dots}{r1,v}
      \fmf{dots}{v,r2}
  \end{fmfgraph}
}}
~,\qquad
\raisebox{-7.5pt}{
\parbox{30pt}{
  \begin{fmfgraph}(30,20)
    \fmfleft{l}
      \fmfright{r1,r2}
      \fmf{gluon}{l,v}
      \fmf{gluon}{r1,v}
      \fmf{gluon}{v,r2}
  \end{fmfgraph}
}}
~,\qquad
\raisebox{-7.5pt}{
\parbox{30pt}{
  \begin{fmfgraph}(30,20)
    \fmfleft{l1,l2}
      \fmfright{r1,r2}
      \fmf{gluon}{l1,v}
      \fmf{gluon}{l2,v}
      \fmf{gluon}{r1,v}
      \fmf{gluon}{v,r2}
  \end{fmfgraph}
}} ~.
\end{align*}
Corresponding to these edges and vertices, we define the following 7 1PI Green's functions:
\begin{gather*}
G^e = 1 - \sum_{\Gamma^e} \frac{\Gamma}{\Sym(\Gamma)};\qquad
G^v = 1 + \sum_{\Gamma^v} \frac{\Gamma}{\Sym(\Gamma)}.
\end{gather*}

In \cite{Sui07} we have shown that the Slavnov--Taylor identities define an ideal in the Hopf algebra $H$ of QCD Feynman graphs. More precisely, the coproduct is compatible with the following identities,
\begin{equation}
\label{ST}
\begin{aligned}
G^\gluc G^\quaglu - G^\gluq G^\qua=0; \\
G^\gluc G^\ghoglu - G^\gluq G^\gho=0;\\
G^\gluc G^\gluc - G^\gluq G^\glu=0.
\end{aligned}
\end{equation}
Hence, the quotient $\tilde H$ of $H$ by this ideal is still a Hopf algebra. In establishing a Hopf subalgebra for QCD, it is essential to work with $\tilde H$ instead of $H$. 
Indeed, the Slavnov--Taylor identities are a crucial ingredient for a closed form of the coproduct on Green's functions. 

\bigskip

Unfortunately, there is no simple expression for the number of insertion places $\ins{\Gamma}{\gamma}$ in QCD. This is due to the fact that there are many different vertices. Nevertheless, there are the following relations between the numbers of vertices, lines and loop number of a fixed graph $\Gamma$ \cite[Lemma 22]{Sui07}
\begin{equation*}
\begin{aligned}
I - V + 1 &=L; \quad &{\rm (a)}
 \qquad & \qquad V_{3F}=I_F + \half E_F; \quad &{\rm (b)}\\
V_3 + 
2 V_4 - E + 2 &= 2L; \quad &{\rm (c)}   
\qquad &  \qquad V_{3G}=I_G + \half E_G. \quad &{\rm (d)}
\end{aligned}
\end{equation*}
The notation is as follows: 
\begin{align*}
I&=I_F + I_G + I_{YM} = \text{ number of internal quark, ghost and gluon lines }\\
E&=E_F + E_G + E_{YM} = \text{ number of external quark, ghost and gluon lines }\\
V&= V_3 + V_4 \\
&= V_{3F} + V_{3G} + V_{3YM} + V_4 = \text{ number of quark-, ghost-, cubic and quartic gluon vertices }
\end{align*}
We can use these expressions to simplify the coproduct on the Green's function corresponding to the vertex/edge $r$. Indeed, as in the previous sections, one can rewrite formula \eqref{cop-green} as
\begin{align}
\label{cop-qcd-pre}
\Delta(G^r) &= \sum_{L=0}^\infty \sum_{ \begin{smallmatrix} \res(\Gamma)=r \\ L(\Gamma)=L \end{smallmatrix}} 
\frac {\left( G^\quaglu  \right)^{V_{3F} }\left( G^\ghoglu \right)^{V_{3G}} 
\left( G^\gluc \right)^{V_{3YM}}  \left( G^\gluq \right)^{V_4} } 
{\left( G^\qua \right)^{I_F} \left( G^\gho \right)^{I_G} \left( G^\glu \right)^{I_{YM} } }
 \otimes \frac{\Gamma}{\Sym(\Gamma)}\\ \nonumber
&=\left( G^\qua \right)^{ \half E_F } \left( G^\gho \right)^{\half E_G} 
 \sum_{L=0}^\infty \sum_\Gamma
\left[ \frac{ G^\quaglu }{ G^\qua } \right]^{V_{3F} }
\left[ \frac{ G^\ghoglu }{ G^\gho } \right]^{V_{3G} }
\frac{ \left( G^\gluc \right)^{V_{3YM}} \left( G^\gluq \right)^{V_4} }{ \left( G^\glu \right)^{I_{YM}} }
\otimes \frac{\Gamma}{\Sym(\Gamma)}
\end{align}
where in going to the second line, we have applied the above equation {\it (b)} and {\it (d)}. We have also understood the notation $E=E(\Gamma), E_F= E_F(\Gamma), \ldots$
We now insert the three Slavnov--Taylor identities in the following form:
\begin{gather*}
\frac{ G^\quaglu }{ G^\qua } = \frac{ G^\gluq }{ G^\gluc }; \qquad
\frac{ G^\ghoglu }{ G^\gho } = \frac{ G^\gluq }{ G^\gluc }; \qquad
\frac{ \left( G^\gluc\right)^2 }{ G^\glu } = G^\gluq,
\end{gather*}
and express everything in terms of the quartic gluon vertex function and gluon propagator. If we then apply the relations {\it (a)} and {\it (c)}, we finally obtain
\begin{align}
\label{cop-qcd}
\Delta(G^r) =
( G^\qua )^{\half E_F} ( G^\gho )^{\half E_G} ( G^\glu )^{\half E_{YM}}  
 \sum_{L=0}^\infty \left[ \frac{ \sqrt{ G^\gluq } }{ G^\glu } \right]^{2L + E - 2} \otimes G_L^r.
\end{align}
Of course, the coefficients $E, E_F, \ldots$ are completely determined by the vertex/edge $r$; together with the factor $\half$ they are precisely what one would expect from wave function renormalization. In the next subsection, we will see that the above equation allows us to derive the well-known relations between unrenormalized, renormalized amplitudes and counterterms in QCD. 

\begin{rem}
The above argument also allows us to re-derive compatibility of the Slavnov--Taylor identities with the coproduct. In fact, the ideal $I$ generated by the left hand sides of Eq. \eqref{ST} defines a Hopf ideal. For this, observe that if we define $X$ and $Y$ by
\begin{gather*}
X= \frac{ G^\quaglu }{ G^\qua }; \qquad Y =  \frac{ G^\gluq }{ G^\gluc },
\end{gather*}
we can replace $X^n$ (with $n=V_{3F}$ to lighten notation) in Eq. \eqref{cop-qcd-pre} by $Y^n$ after addition of $X^n - Y^n$. Now, by induction it follows that 
\begin{equation}
X^n - Y^n = (X-Y)\textup{Pol}(X,Y)
\end{equation}
which is an element in $I$ and similar arguments apply to the other terms. Thus, at the cost of adding extra terms with elements in $I$ on the first leg of the tensor product, one obtains the above formula \eqref{cop-qcd}. When applied to the generators of $I$, one then easily obtains that $\Delta(I) \subset I \otimes H + H \otimes I$.  
\end{rem}

\subsubsection{Renormalized amplitudes and counterterms}
Once again, the QCD Feynman rules induces a map $U$ from $H$ to the algebra of functions in the regularization parameter. We extend this map linearly and obtain the following self-energy functions:
\begin{align*}
\Sigma(g) = U\left( G^\qua \right), \quad
\tilde\Pi(g) =  U\left( G^\gho \right), \quad
\Pi^{\mu\nu} (g)  =  U\left( G^\glu \right), 
\end{align*}
for the quark, ghost and gluon, respectively, as well as the three proper vertex functions:
\begin{align*}
\Gamma^\mu(g) =  U\left( G^\quaglu \right), \quad
G^\mu(g) = U \left( G^\ghoglu \right), \quad
\Gamma^{\mu\nu\sigma}(g) =  U \left( G^\gluc \right), \quad
\Gamma^{\mu\nu\sigma\rho} (g) =  U \left( G^\gluq \right).
\end{align*}
We have adopted the notation of \cite{PT84} and explicitly indicated the dependence on the strong coupling constant $g$. Again, due to the Slavnov--Taylor identities between the above self-energy and vertex functions, the map $U$ vanishes on the ideal $I$ and thus factorizes over $I$ to give a map on $\tilde H$. Moreover, the renormalized self-energy and proper vertex functions are obtained by adding a subscript `$\ren$' on the lhs and replacing $U$ by $R$ on the rhs of the above equations.

The renormalization constants are defined in terms of the counterterm map $C$ of Eq. \eqref{counterterm}:
\begin{gather*}
Z_{1F} = C \left( G^\quaglu \right), \qquad
\tilde Z_1 = C \left( G^\ghoglu \right), \qquad
Z_{1YM} =C \left( G^\gluc \right), \\
\\
Z_{2F} = C \left( G^\qua \right), \qquad
Z_{3YM} =C \left( G^\glu \right), \qquad
\tilde Z_3 = C \left( G^\gho \right), \qquad
Z_{5} = C \left( G^\gluq \right).
\end{gather*}
Since $C$ is an algebra map from $\tilde H$ to functions on the regularization parameter, it vanishes on the ideal $I$. Hence, we deduce the well-known {\it Slavnov--Taylor identities} between the renormalization constants (cf. for instance Eq. (III.59) in \cite{PT84}):
\begin{align*}
\frac{Z_{3YM}}{Z_{1YM}} = \frac{\tilde Z_3}{\tilde Z_1} 
= \frac{Z_{2F}}{Z_{1F}} = \frac{Z_{1YM}} {Z_5}.
\end{align*}

\medskip

Let us now apply $R= C \ast U$ to Equation \eqref{cop-qcd} to derive the well-known formula relating the renormalized and unrenormalized self-energy and vertex functions. First, recall the following formulas (cf. \cite[Eq. (III.55)]{PT84}) for the bare coupling constants
\begin{gather*}
g_{0F} = Z_{1F} Z_{3YM}^{-1/2} Z_{2F}^{-1} g, \qquad
\tilde g_0 = \tilde Z_1 \tilde Z_5^{-1} Z_{3YM}^{-1/2} g, \\
g_{0YM} = Z_{1YM} Z_{3YM}^{-3/2} g, \qquad
g_{05} = Z_5^{1/2} Z_{3YM}^{-1} g,
\end{gather*}
corresponding to the quark-gluon and ghost-gluon interaction and the cubic and quartic gluon self-interaction. We then obtain from Eq. \eqref{cop-qcd}
\begin{gather}
\Sigma_\ren(g) = Z_{2F} \Sigma( g_0 ), \qquad
\tilde \Pi_\ren (g) = \tilde Z_3 \tilde \Pi( g_0 ), \qquad
\Pi^{\mu\nu}_\ren (g) = Z_{3YM} \Pi^{\mu\nu}(g_0),  \nonumber \\
\Gamma^\mu_\ren (g) = Z_{2F} Z_{3YM}^{1/2} \Gamma^\mu(g_0) ,\qquad
G^\mu_\ren (g) =  \tilde Z_{3} Z_{3YM}^{1/2} \Gamma^\mu(g_0) ,
\label{green-qcd}
\\
\Gamma^{\mu\nu\sigma}_\ren (g) = Z_{3YM}^{3/2} \Gamma^{\mu\nu\sigma} (g_0) , \qquad
\Gamma^{\mu\nu\sigma\rho}_\ren (g) =  Z_{3ym}^{2}  \Gamma^{\mu\nu\sigma\rho} (g_0). \nonumber
\end{gather}
Here the argument $g_0$ on the rhs indicates that the regularized functions are computed using the Feynman rules involving the bare coupling constants $g_{0F}, \tilde g_0, g_{0YM}$ and $g_{05}$. That the factors of $\sqrt{Z_5}/Z_{3YM}$ can indeed be absorbed in the bare coupling constants follows from the fact that due to the above Equation {\it (c)}, the power of $g$ that appear in the Green's function at loop order $L$ is precisely $2L+E-2$.

\section*{Acknowledgements}
The author would like to thank ESI in Vienna for its hospitality during the Program `Mathematical and Physical Aspects of Perturbative Approaches to Quantum Field Theory'. Alessandra Frabetti is thanked for pointing out the possibility of obtaining Dyson's formula from the closed formula of the coproduct on Green's functions.

\appendix
\section{Hopf algebras}
For convenience, let us briefly recall the definition of a (commutative) Hopf algebra. It is the dual object to a group and, in fact, there is a one-to-one correspondence between groups and commutative Hopf algebras. 

Let $G$ be a group with product, inverse and identity element. We consider the algebra of representative functions $H = \F(G)$. This class of functions is such that $\F(G \times G) \simeq \F(G) \otimes \F(G)$. For instance, if $G$ is a (complex) matrix group, then $\F(G)$ could be the algebra generated by the coordinate functions $x_{ij}$ so that $x_{ij}(g) = g_{ij} \in \C$ are just the $(i,j)$'th entries of the matrix $g$. 

Let us see what happens with the product, inverse and identity of the group on the level of the algebra $H=\F(G)$. The multiplication of the group can be seen as a map $G \times G \to G$, given by $(g,h) \to gh$. Since dualization reverses arrows, this becomes a map $\Delta: H \to H \otimes H$ called the {\it coproduct} and given for $f \in H$ by
$$
\Delta(f)(g,h) = f(gh).
$$
The property of associativity on $G$ becomes {\it coassociativity} on $H$:
\begin{align} \label{def:Hopf1}
\tag{A1}
(\Delta \otimes \id) \circ \Delta = (\id \otimes \Delta)\circ \Delta,
\end{align}
stating simplfy that $f\big((gh)k\big)= f\big(g(hk)\big)$.

The unit $e \in G$ gives rise to a {\it counit}, as a map $\epsilon: H \to \C$, given by
$\epsilon(f)=f(e)$ and the property $eg=ge=g$ becomes on the algebra level
\begin{equation} \label{def:Hopf2}
\tag{A2}
(\id \otimes \epsilon)\circ \Delta=\id = (\epsilon\otimes\id)\circ\Delta,
\end{equation}
which reads explicitly $f(ge)=f(eg)=f(g)$.

The inverse map $g \mapsto g^{-1}$, becomes the {\it antipode} $S:H\to H$, defined by $S(f)(g)= f(g^{-1})$. The property $g g^{-1}=g^{-1} g= e$, becomes on the algebra level:
\begin{equation} \label{def:Hopf3}
\tag{A3}
m(S\otimes \id)\circ\Delta= m(\id \otimes S)\circ\Delta= 1_H \epsilon,
\end{equation}
where $m: H \otimes H \to H$ denotes pointwise multiplication of functions in $H$.

From this example, we can now abstract the conditions that define a general Hopf algebra. 
\begin{defn}
A {\rm Hopf algebra} $H$ is an algebra $H$, together with two algebra maps $\Delta: H \otimes H \to H$ (coproduct), $\epsilon :H\to \C$ (counit), and a bijective $\C$-linear map $S:H \to H$ (antipode), such that equations \eqref{def:Hopf1}--\eqref{def:Hopf3} are satisfied.
\end{defn}

If the Hopf algebra $H$ is commutative, we can conversely construct a (complex) group from it as follows. Consider the collection $G$ of multiplicative linear maps from $H$ to $\C$. We will show that $G$ is a group. Indeed, we have the {\it convolution product} between two such maps $\phi,\psi$ defined as the dual of the coproduct:
$
(\phi \ast \psi) (X) = (\phi \otimes \psi) (\Delta(X))
$
for $X \in H$. One can easily check that coassociativity of the coproduct (Eq. \eqref{def:Hopf1}) implies associativity of the convolution product: $(\phi \ast \psi) \ast \chi = \phi \ast (\psi \ast \chi)$.
Naturally, the counit defines the unit $e$ by $e(X) = \epsilon(X)$. Clearly $e \ast \phi = \phi = \phi \ast e$ follows at once from Eq. \eqref{def:Hopf2}. Finally, the inverse is constructed from the antipode by setting $\phi^{-1}(X) = \phi(S(X))$ for which the relations $\phi^{-1} \ast \phi = \phi \ast \phi^{-1} = e$ follow directly from Equation \eqref{def:Hopf3}.

With the above explicit correspondence between groups and commutative Hopf algebras, one can translate practically all concepts in group theory to Hopf algebras. For instance, a subgroup $G' \subset G$ corresponds to a {\it Hopf ideal} $I \subset \F(G)$ in that $\F(G') \simeq \F(G)/I$ and viceversa. The conditions for being a subgroup can then be translated to give 
the following three conditions defining a Hopf ideal $I$ in a commutative Hopf algebra $H$
\begin{gather*}
\Delta(I) \subset I \otimes H + H \otimes I,\qquad
\epsilon(I) = 0, \qquad
S(I) \subset I.
\end{gather*}

%\bibliography{references}

\begin{thebibliography}{99}

\bibitem{BK05}
C.~Bergbauer and D.~Kreimer.
\newblock Hopf algebras in renormalization theory: {L}ocality and
  {D}yson-{S}chwinger equations from {H}ochschild cohomology.
\newblock {\em IRMA Lect. Math. Theor. Phys.} 10 (2006)  133--164.

\bibitem{BF00}
C.~Brouder and A.~Frabetti.
\newblock Noncommutative renormalization for massless qed, hep-th/0011161.

\bibitem{BF03}
C.~Brouder and A.~Frabetti.
\newblock {QED} {H}opf algebras on planar binary trees.
\newblock {\em J. Algebra} 267 (2003)  298--322.

\bibitem{BFK06}
C.~Brouder, A.~Frabetti, and C.~Krattenthaler.
\newblock Non-commutative {H}opf algebra of formal diffeomorphisms.
\newblock {\em Adv. Math.} 200 (2006)  479--524.

\bibitem{Col84}
J.~Collins.
\newblock {\em Renormalization}.
\newblock Cambridge University Press, 1984.

\bibitem{CK99}
A.~Connes and D.~Kreimer.
\newblock Renormalization in quantum field theory and the {R}iemann- {H}ilbert
  problem. {I}: {T}he {H}opf algebra structure of graphs and the main theorem.
\newblock {\em Comm. Math. Phys.} 210 (2000)  249--273.

\bibitem{Dys49}
F.~J. Dyson.
\newblock The {S} matrix in quantum electrodynamics.
\newblock {\em Phys. Rev.} 75 (1949)  1736--1755.

\bibitem{EG07}
K.~Ebrahimi-Fard and L.~Guo.
\newblock Rota-{B}axter algebras in renormalization of perturbative quantum
  field theory.
\newblock {\em Fields Inst. Commun.} 50 (2007)  47--105.

\bibitem{IZ1980}
C.~Itzykson and J.~B. Zuber.
\newblock {\em Quantum field theory}.
\newblock McGraw-Hill International Book Co., New York, 1980.
\newblock International Series in Pure and Applied Physics.

\bibitem{Kre98}
D.~Kreimer.
\newblock On the {H}opf algebra structure of perturbative quantum field
  theories.
\newblock {\em Adv. Theor. Math. Phys.} 2 (1998)  303--334.

\bibitem{Kre06}
D.~Kreimer.
\newblock Dyson--{S}chwinger equations: From {H}opf algebras to number theory.
\newblock hep-th/0609004.

\bibitem{Kre05}
D.~Kreimer.
\newblock Anatomy of a gauge theory.
\newblock {\em Ann. Phys.} 321 (2006)  2757--2781.

\bibitem{PT84}
P.~Pascual and R.~Tarrach.
\newblock {QCD}: {R}enormalization for the practitioner.
\newblock In {\em Lecture Notes in Physics}, volume 194. Springer-Verlag,
  Boston, 1984.

\bibitem{Sui06}
W.~D. van Suijlekom.
\newblock The {H}opf algebra of {F}eynman graphs in {QED}.
\newblock {\em Lett. Math. Phys.} 77 (2006)  265--281.

\bibitem{Sui07}
W.~D. van Suijlekom.
\newblock Renormalization of gauge fields: {A} {H}opf algebra approach.
\newblock To appear in {\em Commun. Math. Phys.}

\bibitem{War50}
J.~C. Ward.
\newblock An identity in quantum electrodynamics.
\newblock {\em Phys. Rev.} 78 (1950)  182.

\end{thebibliography}
%\newcommand{\noopsort}[1]{}

\end{fmffile}
\end{document}